\documentclass[12pt]{iopart}

\begin{document}

\title[Levi-Civita transformation]{Connection between Coulomb and harmonic oscillator potentials in relativistic quantum mechanics}

\author{Bo \ Fu, Fu-Lin \ Zhang, Jing-Ling \ Chen$^*$}

\address{Theoretical Physics Division, Chern Institute of
Mathematics, Nankai University, Tianjin, 300071, P.R.China}
\ead{chenjl@nankai.edu.cn}
\begin{abstract}
The Levi-Civita transformation is applied in the two-dimensional
(2D) Dirac and Klein-Gordon (KG) equations with equal external
scalar and vector potentials. The Coulomb and harmonic oscillator
problems are connected via the Levi-Civita transformation. These
connections lead to an approach to solve the Coulomb problems using
the results of the harmonic oscillator potential in the
above-mentioned relativistic systems.

\end{abstract}


\bibliographystyle{unsrt}
\maketitle
\section{Introduction}\label{intro}
Hydrogen atom and harmonic oscillator are usually given in textbooks
as two of the several solvable problems in both classical and
quantum physics \cite{GreinerIn}.The former, whose potential term
takes the Coulomb form, is regarded as a fundamental prototype to
investigate for some more complicated many-electron atoms. And the
latter one is of great importance owing to the fact that many
complicated potentials can approximate to a harmonic oscillator in
the vicinity of their equilibrium points.

There is a close relationship between the two fundamental models
\cite{ChenA}. This topic has attracted considerable interest, partly
because these simple systems concern concepts such as accidental
degeneracy, dynamical symmetry, etc. and also because the
relationship between them is of practical value, for example in the
computation of matrix elements \cite{Wolf}, in the derivation of the
Coulomb path integral \cite{Path}, and in the construction of
coherent states \cite{Coherent}. A convenient method to establish
the connection between hydrogen atom and harmonic oscillator is the
Kustaanheimo-Stiefel transformation \cite{Stiefel,Ikeda,Kibler},
which arises from the issue on the relation between the Kepler
problem and the classical oscillator in celestial mechanics. It is
obtained by generalizing the Levi-Civita transformation
\cite{levi-civita,Antonyan} which works only for the two dimensional
cases. This method could be naturally applied to transform the
Schr$\ddot{o}$dinger equation of hydrogen atom to that of harmonic
oscillator in non-relativistic quantum mechanics. Only the
Levi-Civita transformation would be taken into account in the rest
sections of the paper since we shall discuss the problems only in
two dimensional space.

It is well-known that the non-relativistic quantum mechanics is an
approximate theory of the relativistic one. In relativistic quantum
mechanics, the motion of spin-$0$ and spin-$1/2$ particles satisfies
the KG and the Dirac equations respectively.

The Kustaanheimo-Stiefel transformation and Levi-Civita
transformation are in fact related to the well-known dynamical
symmetries of the two non-relativistic models
\cite{GreinerSym,Gross}. For instance, the conserved quantities of
2D hydrogen atom generate the Lie group $SO(3)$ , whereas that of 2D
harmonic oscillator generate the Lie group $SU(2)$ \cite{Wybourne}.
The correspondence between the motion equations of the two models is
to a large extent decided by the relation between the two Lie groups
\cite{louck}.Recently, the dynamical symmetries both in the KG and
the Dirac equations have been reported. Namely, in the KG and the
Dirac systems, the Hamiltonian with equal scalar and vector Coulomb
or harmonic oscillator potentials have the same dynamical symmetries
as their non-relativistic counterparts
\cite{Ginocchio,ZhangDirac,ZhangKG}. The discussion above suggests
that there should be a coordinate transformation connecting the
relativistic systems with the $SO(3)$ and the $SU(2)$ dynamical
symmetries. In this paper, we will show that it is nothing but the
Levi-Civita transformation.

For the sake of brevity, the spin-$0$ (or $1/2$) particle in equal
scalar and vector Coulomb or harmonic oscillator potentials are
called KG (or Dirac) hydrogen atom or harmonic oscillator
respectively. The relativistic units are chosen in this paper as
$c=\hbar=1$. Before we go further, in Sec. \ref{review}, we would
first make a brief review for the Levi-Civita transformation in the
non-relativistic quantum mechanics \cite{Antonyan}. In Sec. \ref{KG}
and Sec. \ref{Dirac} we will show the Levi-Civita transformation in
the KG equation and Dirac equation separately. Conclusion and
discussion will be made in the last section.

\section{Non-relativistic quantum mechanics }\label{review}

We shall start with the time-independent Schr$\ddot{o}$dinger
equation for a 2D hydrogen atom
\begin{eqnarray}\label{Sch}
 H\Psi=E\Psi,\ \ \ H=\frac{\textbf{p}^2}{2\mu}-\frac{\kappa}{r},
\end{eqnarray}
where $\mu$ is the reduced mass of the hydrogen atom, $\kappa =
e^2$, \ $\textbf{p}^2=-\Sigma^2_{i=1}(\partial^2/\partial x^2_i)$,
the $x_i$ being the Cartesian coordinates, and
$r=(x^2_1+x^2_2)^{1/2}$. We now begin to transform the problem into
a 2D harmonic oscillator via the Levi-Civita transformation.
Introducing the variables $u_1$ and $u_2$, the  transformation can
be written as
\begin{eqnarray}\label{ks}
x_1=u^2_1-u^2_2,\ \ \ x_2=2u_1u_2.
\end{eqnarray}
Under this transformation we have $r=u^2=u^2_1+u^2_2$. The
Schr$\ddot{o}$dinger equation (\ref{Sch}) becomes
\begin{eqnarray}\label{3}
\biggr[-\frac{1}{8\mu}\frac{1}{u^2}\displaystyle{\sum^2_{i=1}}\frac{\partial^2}{\partial
u^2_i}-\frac{\kappa}{r}\biggr]\Psi=E\Psi.
\end{eqnarray}
After multiplying on both sides of the equation (\ref{3}) by $u^2$,
we get
\begin{eqnarray}
\biggr[-\frac{1}{8\mu}\displaystyle{\sum^2_{i=1}}\frac{\partial^2}{\partial
u^2_i}-Eu^2\biggr]\Psi=\kappa\Psi.
\end{eqnarray}
This equation turns into the form of the Schr$\ddot{o}$dinger
equation of the 2D harmonic oscillator after assuming that $E<0$
(for bound motions), and making the definitions
\begin{eqnarray}\label{def}
m=4\mu,\ \ \ \omega=(-E/2\mu)^{1/2},\ \ \ \epsilon=\kappa.
\end{eqnarray}
Then, we obtain
\begin{eqnarray}
\biggr(-\frac{1}{2m}\displaystyle{\sum^2_{i=1}}\frac{\partial^2}{\partial
u^2_i}+\frac{1}{2}m\omega^2u^2\biggr)\Psi=\epsilon\Psi,
\end{eqnarray}
or $\mathcal{H}\Psi=\epsilon\Psi$, with
\begin{eqnarray}
\mathcal{H}=-\frac{1}{2m}\displaystyle{\sum^2_{i=1}}\frac{\partial^2}{\partial
u^2_i}+\frac{1}{2}m\omega^2u^2.
\end{eqnarray}
Here, $\mathcal{H}$ and $\epsilon$ are consistent with the
Hamiltonian and the energy eigenvalue of a 2D harmonic oscillator
respectively. Considering the energy spectrum of the 2D harmonic
oscillator in non-relativistic quantum mechanics, one can obtain
\begin{eqnarray}
\kappa=\epsilon=(n_1+n_2+1)\omega, \ \ \ n_1,n_2=0,1,2...
\end{eqnarray}
From Eq.(\ref{def}), the energy levels of the 2D hydrogen atom could
be derived immediately
\begin{eqnarray}\label{ClassicalHE}
E\equiv E_n=-\frac{\kappa}{2a}\frac{1}{n^2} \ \ \ n=1,2,3...
\end{eqnarray}
where $a=1/(\mu\kappa)$ is the Bohr radius, and
$n=\frac{1}{2}(n_1+n_2+1)$. It is obvious that the value of
$n_1+n_2$ must be odd number to guarantee that $n$ is positive
integer. Hence the expression in Eq. (\ref{ClassicalHE}) is the
familiar Bohr formula for the energy eigenvalues of hydrogen atom.
The correspondence of energy level is not complete, where the energy
level of harmonic oscillator only partly corresponds with that of
hydrogen atom. The cause of this problem will be illustrated in the
following section.

\section{KG equation }\label{KG}

The KG equation with scalar potential $V_S$ and vector potential
$V_V$ is given by
\begin{eqnarray}\label{kg1}
\biggr\{p^2+[M+V_S]^2-\biggr[i\frac{\partial}{\partial
t}-V_V\biggr]^2\biggr\}\Psi=0,
\end{eqnarray}
where $M$ is the mass, $p^2=-\Sigma^2_{i=1}(\partial^2/\partial
x^2_i)$ is the momentum. For the time-independent potentials
$V_S=V_V=V(r)/2$, the KG equation (\ref{kg1}) becomes
\begin{eqnarray}\label{kg2}
\biggr[-\displaystyle{\sum^2_{i=1}}\frac{\partial^2}{\partial
x_i^2}+(M+E)V(r)-(E^2-M^2)\biggr]\Psi=0,
\end{eqnarray}
where $E$ is the relativistic energy. If the potential term $V(r)$
takes the Coulomb form, i.e. $V(r)=-\kappa/r$, we obtain the
eigen-equation of the KG hydrogen atom
\begin{eqnarray}\label{kg3}
\biggr[-\displaystyle{\sum^2_{i=1}}\frac{\partial^2}{\partial
x_i^2}-(M+E)\frac{\kappa}{r}-(E^2-M^2)\biggr]\Psi=0.
\end{eqnarray}
Under the Levi-Civita transformation in Eq. (\ref{ks}), the KG
equation (\ref{kg3}) becomes
\begin{eqnarray}\label{kg4}
\biggr[-\displaystyle{\sum^2_{i=1}}\frac{\partial^2}{\partial
u^2_i}-4(M+E)\kappa-4(E^2-M^2)u^2\biggr]\Psi=0.
\end{eqnarray}
Setting
\begin{eqnarray}\label{def1}
\left\{
  \begin{array}{ll}
    4\kappa=\epsilon-m \\
    M+E=m+\epsilon \\
    -4(E-M)=\frac{1}{2}m\omega^2
  \end{array}
\right. \Rightarrow \left\{
  \begin{array}{ll}
    \kappa=\frac{1}{4}(\epsilon-m) \\
    M+E=m+\epsilon \\
    M-E=\frac{1}{8}m\omega^2
  \end{array},
\right.
\end{eqnarray}
the KG equation (\ref{kg4}) would become
\begin{eqnarray}\label{kg5}
\biggr[-\displaystyle{\sum^2_{i=1}}\frac{\partial^2}{\partial
u_i^2}+\frac{1}{2}m\omega^2u^2(m+\epsilon)-(\epsilon^2-m^2)\biggr]\Psi=0¡£
\end{eqnarray}
It is consistent with the eigen-equation of the 2D KG harmonic
oscillator, with mass $m$, frequency $\omega$ and energy level
$\epsilon$.

The Levi-Civita transformation also reveals the relationship between
the energy eigenvalues of the 2D KG hydrogen atom and 2D KG harmonic
oscillator. The energy spectra of 2D harmonic oscillator is the real
root of the cubic equation \cite{ZhangKG}
\begin{eqnarray}\label{KGOE}
\begin{array}{l}
(\epsilon-m)^2(\epsilon+m)^2=2m\omega^2(\epsilon+m)(n+1)^2 \\
n=2j=0,1,2...
\end{array}
\end{eqnarray}
 Substituting the Eq. (\ref{def1}) into the cubic equation
(\ref{KGOE}), one can obtain
\begin{eqnarray}\label{KGHA}
E=\frac{\pm s^2-\kappa^2}{s^2+\kappa^2}M, \ \ \ s=2j+1=1,2,3...
\end{eqnarray}
It takes the same form as the energy level of the KG hydrogen atom
but differs in the values of n \cite{ZhangKG}. In the energy level
of the KG hydrogen atom, the value of $n$ is a positive odd number,
i.e. $n=2j+1=1,3,5...$ Hence the correspondence of the energy levels
of the two systems is not complete. This problem results from the
value of $j$ in the expression of energy spectra. From
\cite{ZhangKG} $j(j+1)$ is the eigenvalue of the Casimir operator.
We introduce the orbit angular momentum $l_h$ and $l_o$ for 2D KG
hydrogen atom and harmonic oscillator respectively,
\begin{eqnarray} \label{LHLO}
l_h=\frac{1}{i}x_1\frac{\partial}{\partial
x_2}-\frac{1}{i}x_2\frac{\partial}{\partial x_1}, \ \ \
l_o=\frac{1}{i}u_1\frac{\partial}{\partial
u_2}-\frac{1}{i}u_2\frac{\partial}{\partial u_1},
\end{eqnarray}
which are the conserved quantities of the two systems and whose
eigenvalue is integer. The relationship between this two conserved
quantities could be established via the Levi-Civita transformation.
Combining the definition of the angular momentum with the
Levi-Civita transformation (\ref{ks}), we could get the relation
\begin{eqnarray} \label{RoL}
l_h=\frac{l_o}{2}
\end{eqnarray}
 The $l_h$ and $l_o/2$ are just the
normalized generators of the two systems. The relation of the orbit
angular momentum between the two systems is the same as the
relations of the other normalized generators of the KG hydrogen atom
and the KG harmonic oscillator. Hence the value of $j$ in KG
hydrogen atom is non-negative integer whereas the value of $j$ in KG
harmonic oscillator is non-negative integer and half-odd-integer,
which leads to the incomplete correspondence of the energy spectrum
of the KG hydrogen atom and the KG harmonic oscillator. The
illustration for the controversy of the correspondence of energy
level of the two KG systems could also be applied to the
inconsistency in Section \ref{review} and Section \ref{Dirac}.

\section{Dirac equation }\label{Dirac}

The Hamiltonian of the Dirac hydrogen atom with equal scalar and
vector Coulomb potentials is given by
\begin{eqnarray}
\begin{array}{lll}
H&=&\alpha_1 p_1 + \alpha_2 p_2+\beta M+(1+\beta)\frac{V(r)}{2},\ \  \\
V(r)&=&-\frac{\kappa}{r} ,
\end{array}
\end{eqnarray}
where $\alpha_1=\biggr( \begin{array}{cc} 0 & 1 \\ 1 & 0
\end{array} \biggr)$,
$\alpha_2=\biggr( \begin{array}{cc} 0&-i \\ i&0 \end{array}
\biggr)$, $\beta=\biggr( \begin{array}{cc} 1&0 \\ 0&-1 \end{array}
\biggr)$ are the Pauli matrices, $(p_1, p_2)$ are the 2D linear
momenta, $(x_1, x_2)$ are the spatial coordinates with magnitude
$r$, and $M$ is the mass. The Dirac equation can be written as
\begin{eqnarray}\label{Diraceq}
H\Psi=E\Psi, \ \ \ H=\biggr( \begin{array}{cc}
M-\frac{\kappa}{r}&p_{1x}-ip_{2x} \cr p_{1x}+ip_{2x}& -M
\end{array} \biggr).
\end{eqnarray}
We could get the relation of linear momentum via the Levi-Civita
transformation as
\begin{eqnarray}\label{rela1}
p_{1u}-ip_{2u}&=&2(u_1+iu_2)(p_{1x}-ip_{2x}), \nonumber \\
p_{1u}+ip_{2u}&=&2(p_{1x}+ip_{2x})(u_1-iu_2),
\end{eqnarray}
where $p_{ix}$ is the 2D linear momentum of hydrogen atom, $p_{iu}$
is the 2D linear momentum of harmonic oscillator. In order to give
an obvious result, we introduce the following transformation into
the Dirac equation (\ref{Diraceq})
\begin{eqnarray}\label{diractrans}
C(H-E)C^{\dag}D\Psi=0,
\end{eqnarray}
where $C=\biggr( \begin{array}{cc} 2\tau&0 \cr 0&1  \end{array}
\biggr)$, $C^{\dag}=\biggr( \begin{array}{cc} 2\tau^{\dag}&0 \cr 0&1
\end{array} \biggr)$, $D=\biggr( \begin{array}{cc} \frac{\tau}{2u^2}&0 \cr 0&1 \end{array} \biggr)$,
$\tau=u_1+iu_2$, $\tau^{\dag}=u_1-iu_2$, and $\Psi=\biggr(
\begin{array}{cc}  \Psi_1 \cr
\Psi_2  \end{array} \biggr)$ is the eigenfunction of the Hamiltonian
for relativistic hydrogen atom. Then, the Dirac equation Eq.
(\ref{Diraceq}) becomes
\begin{eqnarray}
\biggr( \begin{array}{cc} 4Mu^2-4\kappa-4Eu^2&p_{1u}-ip_{2u} \cr
p_{1u}+ip_{2u}&-M-E\end{array} \biggr) \biggr(
\begin{array}{cc}\frac{u_1+iu_2}{2u^2}\Psi_1 \cr \Psi_2
\end{array} \biggr)=0.
\end{eqnarray}
From Eq.(\ref{def1}) the Dirac equation (\ref{Diraceq}) would become
\begin{eqnarray}\label{oeq}
(\mathcal{H}-\epsilon)\Phi=0, \ \ \mathcal{H}=\biggr(
\begin{array}{cc}
m+\frac{1}{2}m\omega^2u^2 & p_{1u}-ip_{2u} \cr p_{1u}+ip_{2u} &
-m\end{array} \biggr),
\end{eqnarray}
where $\Phi=\biggr( \begin{array}{cc}\Phi_1 \cr \Phi_2  \end{array}
\biggr) = \biggr( \begin{array}{cc} \frac{\tau}{2u^2}\Psi_1 \cr
\Psi_2  \end{array} \biggr) $. The Eq. (\ref{oeq}) is consistent
with the Dirac equation for Dirac harmonic oscillator
\cite{ZhangDirac}, where $\mathcal{H}$ could be seen as the
Hamiltonian of the Dirac harmonic oscillator and the wave function
$\Phi$ could be considered as the eigenfunction of the Hamiltonian
$\mathcal{H}$.

Following the same procedure as Section \ref{KG}, we could also get
the energy levels of the 2D Dirac hydrogen atom from the result of
2D Dirac harmonic oscillator. The energy spectrum of the Dirac
harmonic oscillator given in \cite{ZhangDirac} can be expressed as
the real roots of the following equation,
\begin{eqnarray}\label{HOlevel1}
\frac{(\epsilon-m)^2(\epsilon+m)^2}{2m\omega^2}-(\epsilon+m)(n+1)^2=0,
\ \ \
\end{eqnarray}
where $n=2j=0,1,2...$ Substituting the definitions (\ref{def}) into
the energy spectrum expression of the 2D Dirac harmonic oscillator,
we could obtain the following expression
\begin{eqnarray}\label{HAlevel}
E=\frac{\pm n^2-\kappa^2}{n^2+\kappa^2}M, \ \ \ n=2j+1=1,3,5...
\end{eqnarray}
which is consistent with the energy eigenvalue of the 2D Dirac
hydrogen atom \cite{ZhangDirac}, where $E$ and $M$ could be
considered as the relativistic energy and mass of the 2D hydrogen
atom. Only a part of the energy level of the 2D Dirac harmonic
oscillator corresponds to that of the 2D hydrogen atom. The main
reason is the same as is given in Section \ref{KG}. The conserved
angular momenta of the 2D Dirac hydrogen atom and harmonic
oscillator are given by \cite{ZhangDirac},
\begin{eqnarray}
L_o=\biggr( \begin{array}{cc} l_o&0 \cr 0& U^{\dag}_{px} l_o
U_{px}\end{array} \biggr), \ \  L_h=\biggr( \begin{array}{cc}l_o&0
\cr 0& U^{\dag}_{pu} l_o U_{pu}
\end{array} \biggr)
\end{eqnarray}
where $U_{px}=(p_{1x}- i p_{2x})/ \sqrt{p^2_{1x}+ p^2_{2x}} $ ,
$U_{pu}=(p_{1u}- i p_{2u})/\sqrt{p^2_{1u}+ p^2_{2u}} $,  $l_o$ and $
l_h$ are defined in Eq. (\ref{LHLO}).
 Their relation under the Levi-Civita
transformation can be easily derived as
\begin{eqnarray}
\biggr( \begin{array}{cc}  1&0 \cr 0&U_u\end{array} \biggr) L_h
\biggr( \begin{array}{cc}  1&0 \cr 0&U^{\dag}_u\end{array}
\biggr)=\frac{L_o}{2}
\end{eqnarray}
where $U_u=(u_1 -i u_2)/u$ and $U^{\dag}_u=(u_1 +i u_2)/u$. This
explains the discrepancy of the quantum numbers in the 2D Dirac
hydrogen atom and 2D Dirac harmonic oscillator.

\section{Conclusion and discussion}\label{conclu}

The Levi-Civita transformation connects the 2D Coulomb problem and
2D harmonic oscillator in non-relativistic quantum mechanics. It is
in close touch with the dynamical symmetries of these two models.
Both the spin-$0$ and spin-$1/2$ particles in equal scalar and
vector Coulomb or harmonic oscillator potentials have the $SO(3)$ or
$SU(2)$ dynamical symmetries. In this paper, we have shown that the
Levi-Civita transformation can be applied to transform the 2D KG (or
Dirac) hydrogen atom to KG (or Dirac) harmonic oscillator. Taking
the relation between the conserved angular momenta into account, the
Levi-Civita transformation leads to an approach to solve the KG (or
Dirac) hydrogen atom by using the results of the KG (or Dirac)
harmonic oscillator. In addition, the connection between 3D hydrogen
atom and 4D isotropic harmonic oscillator has also been a subject of
considerable interest in the last three decades
\cite{Kibler,Operator,Kibler3} and the Kustaanheimo-Stiefel
transformation between them in non-relativistic quantum mechanics
has been well-known. Hence we could go further to generalize it to
the relativistic case by a similar procedure.

\section*{Acknowledgments}
This work is supported in part by NSF of China (Grants No. 10975075)
and Program for New Century Excellent Talents in University. The
Project-sponsored by SRF for ROCS, SEM.

\section*{References}

\bibliography{LCT}
\end{document}